\documentclass[aps,pra,reprint,amsmath,amssymb,superscriptaddress,longbibliography,floatfix]{revtex4-1}
\usepackage{graphicx}
\usepackage[colorlinks=true,urlcolor=blue,citecolor=blue,linkcolor=blue,bookmarks=false,pdfstartview={FitH}]{hyperref}

\begin{document}

\title{Lattice dynamics, crystal-field excitations and quadrupolar fluctuations of YbRu$_2$Ge$_2$}

\author{Mai Ye}
\email{mye@physics.rutgers.edu}
\affiliation{Department of Physics and Astronomy, Rutgers University, Piscataway, NJ 08854, USA}
\author{E. W. Rosenberg}
\affiliation{Department of Applied Physics, Stanford University, CA 94305, USA}
\author{I. R. Fisher}
\affiliation{Department of Applied Physics, Stanford University, CA 94305, USA}
\author{G. Blumberg}
\email{girsh@physics.rutgers.edu}
\affiliation{Department of Physics and Astronomy, Rutgers University, Piscataway, NJ 08854, USA}
\affiliation{Laboratory of Chemical Physics, National Institute of Chemical Physics and Biophysics, 12618 Tallinn, Estonia}

\date{\today}

\begin{abstract}
We employ polarization resolved Raman scattering spectroscopy to 
study ferroquadrupolar (FQ) fluctuations and 
crystal-field (CF) excitations in YbRu$_2$Ge$_2$ 
heavy-fermion metal with FQ transition at T$_Q$=10\,K. 
We demonstrate that the electronic static Raman 
susceptibilities in 
quadrupolar symmetry channels 
exhibit nearly Curie law behavior, and that the electron-lattice coupling is essential 
for the FQ transition at T$_Q$. 
We establish the CF level scheme of the Yb$^{3+}$ ground state 
$^2F_{7/2}$ multiplet.
We study the lattice dynamics and demonstrate coupling 
between CF transitions and phonon modes. 
\end{abstract}

\maketitle

\section{Introduction\label{sec:Intro}}

Multipolar interactions and related ordering phenomena have attracted 
great interest because, unlike commonly-known long-range orders of 
electric or magnetic dipole moments, multipoles are often related to 
more exotic phases which are difficult to probe directly by 
conventional methods~\cite{Kuramoto2009,Santini2009}. 
For systematic investigation of the collective behavior of multipole 
moments, $f$-electron systems are suitable choices since the strong 
coupling between spin and orbital degrees of freedom of $f$-electrons 
facilitates multipole formation.  
Indeed, the actinide dioxides with $5f$-electrons exhibit a variety 
of multipolar ordering phenomena~\cite{Santini2009}.  
For lanthanides with $4f$-electrons, multipolar, and especially 
quadrupolar orders, have been discovered for different 
systems~\cite{Tm2000,Dy2000,Np2002,Pr2006,Cameron2016}.                

YbRu$_2$Ge$_2$, a heavy-fermion metal with tetragonal structure 
(space group I4/mmm, No. 139; point group D$_{4h}$), has been 
suggested to hold a ferroquadrupolar (FQ) order at low 
temperature~\cite{Jeevan2006,Takimoto2008,Thalmeier2008,Nicklas2010,Prasad2010,Jeevan2011,Fisher2018,ZXShen2019}. 
It undergoes a second-order phase transition at T$_Q$=10\,K, before 
entering into an antiferromagnetic (AFM) phase below T$_{N1}$=6.5\,K~\cite{Jeevan2006,Jeevan2011}. 
At T$_{N2}$=5.5\,K, there may exist a small change in the magnetic 
structure~\cite{Jeevan2011,Nicklas2010}. Early studies show that the 
transition at T$_Q$ is not magnetic, and T$_Q$ increases when 
magnetic field is applied along the easy 
direction~\cite{Jeevan2006,Jeevan2011}. This behavior at T$_Q$ is 
similar to that of TmAu$_2$ at its FQ ordering 
temperature~\cite{Kosaka1998}, hence suggesting a FQ phase in 
YbRu$_2$Ge$_2$. The existence of a FQ order is further supported by 
recent elastoresistivity studies where above T$_Q$ the 
elastoresistivity in the quadrupolar symmetry channels displays a 
Curie-Weiss behavior~\cite{Fisher2018}. Below T$_Q$, an orthorhombic 
structural distortion is observed by X-ray diffraction, which 
confirms that the FQ state breaks B$_{1g}$ ($x^{2}-y^{2}$) 
symmetry~\cite{Fisher2018}.              

The FQ order, namely the ordering of Yb$^{3+}$ 4f-electron charge 
distribution at zero wavevector, can be probed indirectly by studying 
the lattice dynamics and crystal-field (CF) excitations. In a FQ 
arrangement, aligned charge quadrupoles uniformly distort the lattice 
via a coupling between the quadrupole moment and the strain field 
with the same symmetry. The induced distortion reduces the 
point-group symmetry of the lattice system, splitting degenerate 
phonon modes; the distortion also modifies the energy and lifetime of 
the phonon modes of the same symmetry. Such anomalies can be revealed 
by investigating the phonon spectra. Besides, the quadrupolar moments 
are carried by the CF ground state of Yb$^{3+}$. The tetragonal CF 
potential splits the $^2F_{7/2}$ ground multiplet into two $\Gamma_6$ 
and two $\Gamma_7$ Kramers doublets. 
The magnetic entropy right above T$_Q$ is nearly 
R\,ln\,4~\cite{Jeevan2006}, suggesting that the CF ground state is a 
quasi-quartet consisting of two quasi-degenerate Kramers doublets. 
The quasi-quartet ground state was recently confirmed by 
angle-resolved photo-emission spectroscopy  
studies~\cite{ZXShen2019}. This quasi-quartet near degeneracy is 
essential for forming a quadrupolar ground state and deserves a 
detailed study.        

Raman spectroscopy is a conventional tool for studying phonon modes~\cite{Cardona1982} and CF excitations~\cite{Cardona2000}. Here we study the lattice dynamics, low-energy quadrupole fluctuations, and CF excitations in YbRu$_2$Ge$_2$. We assign four Raman-active phonon modes, and reveal an anomalous intensity enhancement of two phonon modes on cooling. The three CF transitions within the $^2F_{7/2}$ ground multiplet are identified and a CF level scheme is in turn established. We demonstrate that low-energy Raman response undergoes remarkable enhancement on cooling towards T$_Q$ and that the static electronic Raman susceptibility in the corresponding quadrupole channels follows nearly perfect Curie behavior, signifying that the relatively strong coupling to the lattice in the B$_{1g}$-symmetry channel enhances by about 10\,K the vanishingly small electronic Weiss temperature to the FQ transition temperature T$_Q$.

\section{Experimental\label{sec:Exp}}

Single crystals of YbRu$_2$Ge$_2$ were grown by flux method; details of the growth can be found in Ref~\cite{Fisher2018}. Two samples were used in this study: one was cleaved in ambient condition to expose its xy crystallographic plane, the other had a clean as-grown xz crystallographic plane. The xy crystallographic plane was examined under a Nomarski microscope to find about 200$\times$200\,$\mu$m$^{2}$ strain-free area.   

Raman scattering measurements were performed in a quasi-back scattering geometry from sample placed in a continuous helium-gas-flow cryostat. We used 476.2, 647.1 and 752.5\,nm lines from a Kr$^+$ ion laser for excitation. Incident light with no more than 14\,mW power was focused to a 50$\times$100\,$\mu$m$^{2}$ spot. Particularly, for measurements below 10\,K, the power of the incident light was reduced to 2\,mW. The temperatures reported in this paper were corrected for laser heating, which was estimated to be \,0.75\,$\pm$\,0.25\,K/mW~\footnote{Optical absorption coefficient and thermal conductivity data for YbRu$_2$Ge$_2$, which are required for a model estimation of the laser heating~\cite{Maksimov1992}, are currently unavailable. Hence, we base the estimate of laser heating at low temperatures on comparison with other heavy-fermion metals with comparable electrical conductivity~\cite{Fisher2018}: CeB$_{6}$~\cite{Takase1980} and URu$_{2}$Si$_{2}$~\cite{Palstra1986}, for which laser heating for the same experimental set-up was established in the range of 0.5$\sim$1.0\,K/mW.}.                       

Seven polarization configurations were employed to probe excitations in different symmetry channels. The relationship between the scattering geometries and the symmetry channels~\cite{Hayes2004} is given in Table~\ref{tab:Exp1}.

\begin{table}[t]
\caption{\label{tab:Exp1}The relationship between the scattering geometries and the symmetry channels. For scattering geometry E$_{i}$E$_{s}$, E$_{i}$ and E$_{s}$ are the polarizations of incident and scattered light; X, Y, X', Y' and Z are the [100], [010], [110], [1$\overline{1}$0] and [001] crystallographic directions; R and L are right and left circular polarizations. A$_{1g}$, A$_{2g}$, B$_{1g}$, B$_{2g}$ and E$_{g}$ are the irreducible representations of the D$_{4h}$ group.}
\begin{ruledtabular}
\begin{tabular}{ll}
Scattering Geometry&Symmetry Channel\\
\hline
XX&A$_{1g}$+B$_{1g}$\\
XY&A$_{2g}$+B$_{2g}$\\
X'X'&A$_{1g}$+B$_{2g}$\\
X'Y'&A$_{2g}$+B$_{1g}$\\
XZ&E$_{g}$\\
RR&A$_{1g}$+A$_{2g}$\\
RL&B$_{1g}$+B$_{2g}$\\
\end{tabular}
\end{ruledtabular}
\end{table}

We used a custom triple-grating spectrometer with a 
liquid-nitrogen-cooled charge-coupled device (CCD) detector for 
analysis and collection of the scattered light. The data were corrected for the spectral response of the system. The measured secondary-emission intensity $I(\omega,T)$ is related to the Raman response $\chi''(\omega,T)$ by $I(\omega,T)=[1+n(\omega,T)]\chi''(\omega,T)+L(\omega,T)$, where $n$ is the Bose factor, $\omega$ is energy, $T$ is temperature. $L(\omega,T)$ represents the far tail of photo-luminescence, which in the narrow spectral window of interest was approximated by a linear frequency dependence.            

\section{Results and Discussion\label{sec:Res}}

\subsection{Lattice Dynamics\label{subsec:LD}}

\begin{figure}
\includegraphics[width=0.48\textwidth]{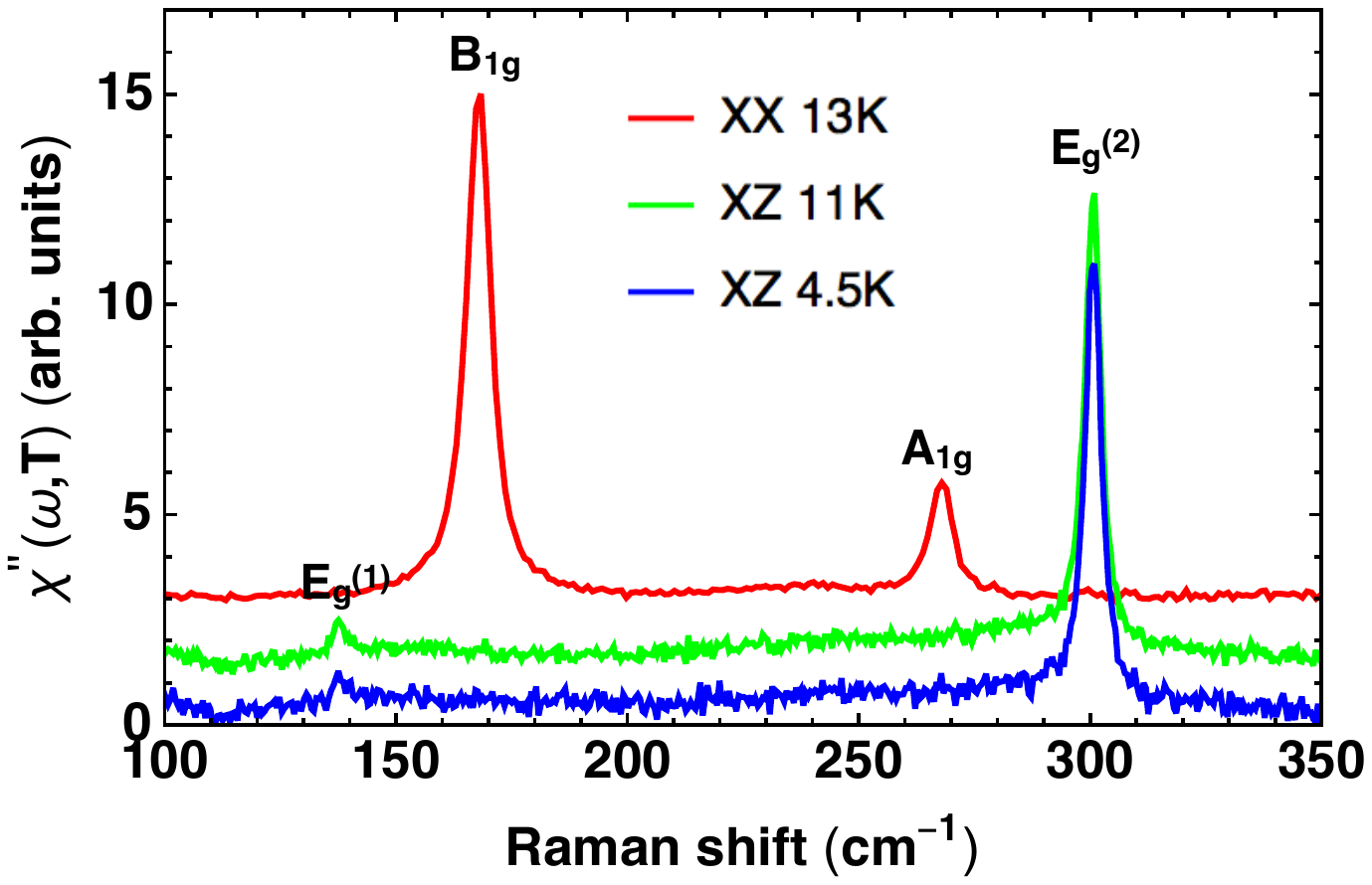}
\caption{\label{fig:P}Raman response $\chi''(\omega,T)$ of four Raman-active optical phonon modes at low temperature measured with the 647.1\,nm excitation. The XX and XZ spectra are offset by 1.5 and 3 arbitrary units (arb. units), respectively. The spectral resolution is 1.0\,cm$^{-1}$.}
\end{figure}

The spectra of phonon modes are presented in Fig.~\ref{fig:P}. By 
group theory, four Raman-active optical phonon modes are expected for 
YbRu$_2$Ge$_2$ structure: $A_{1g}\oplus B_{1g}\oplus 2E_{g}$. 
A$_{1g}$ and B$_{1g}$ modes are accessible in XX geometry and E$_{g}$ modes in XZ geometry. 
The phonon energies at 13\,K are tabulated in Table~\ref{table:ModeSummary}.

\begin{table}[b]
\caption{\label{table:ModeSummary}Summary of the energy of the phonon 
and crystal-field (CF) modes. The coupled CF and phonon modes are labeled by ``(c)". Results of this work are compared to inelastic neutron scattering (INS) study~\cite{Jeevan2010}. Units are cm$^{-1}$.}
\begin{ruledtabular}
\begin{tabular}{lll}
Mode&This work&INS\\
\hline
$\Gamma_6^{(1)}\rightarrow\Gamma_7^{(1)}$&2&--\\
$\Gamma_6^{(1)}\rightarrow\Gamma_7^{(2)}$&95&89\\
$\Gamma_6^{(1)}\rightarrow\Gamma_6^{(2)}$ (c)&239&--\\
\hline
A$_{1g}$ (c)&268&260\\
B$_{1g}$&168&170\\
E$_{g}^{(1)}$&138&--\\
E$_{g}^{(2)}$ (c)&301&--\\
\end{tabular}
\end{ruledtabular}
\end{table}

\begin{figure}
\includegraphics[width=0.44\textwidth]{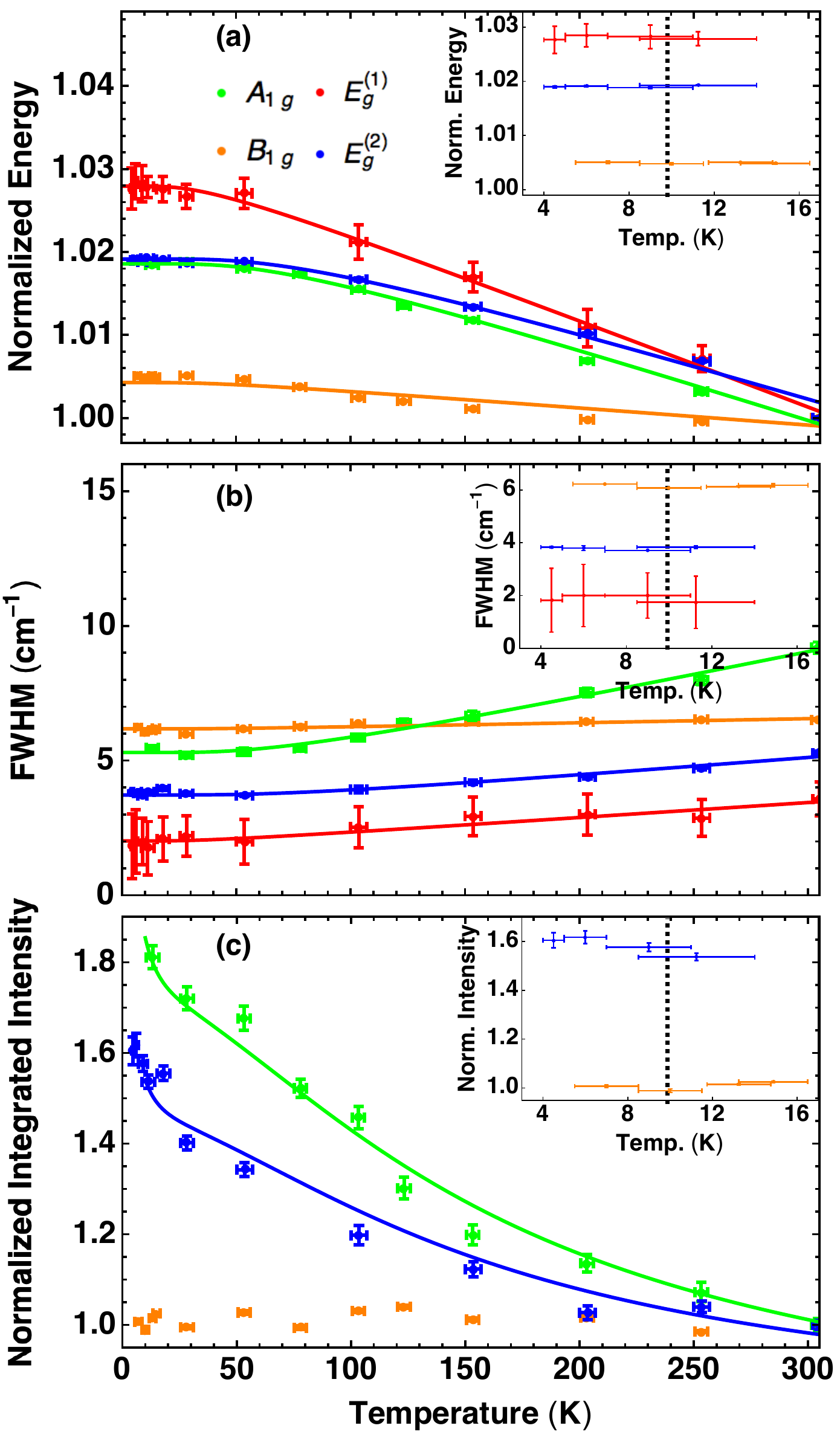}
\caption{\label{fig:P_Para} 
Temperature dependence of (a) the energy, (b) the FWHM and (c) the 
integrated intensity of the Raman-active optical phonon modes. The 
energy and integrated intensity are normalized to their respective 
value at 304\,K. The integrated intensity of the very weak 
E$_{g}^{(1)}$ phonon mode is not shown. The solid lines in (a) and 
(b) represent the fits to anharmonic decay 
model~\cite{Klemens1966,Cardona1984}, while the solid lines in (c) 
represent the fits to Eq.~(\ref{eq:Int}). Insets: zoom-in of the 
low-temperature data points showing how the physical properties 
change across the phase-transition temperature T$_Q$=10\,K. The 
dashed line in the insets indicate T$_Q$. The vertical error bars are derived from Lorentzian fits and represent one standard deviation; the horizontal error bars are derived from the uncertainty of laser heating estimation.}
\end{figure}

In Fig.~\ref{fig:P_Para} we show the temperature dependence of the spectral parameters (energy, FWHM, and integrated intensity) of the phonon modes. The spectral parameters were obtained by fitting the measured spectral peaks with Lorentzian lineshapes.

The temperature dependence of both frequency and FWHM of the phonon 
modes [Fig.~\ref{fig:P_Para}(a) and (b)] is in accordance with a 
simple model assuming anharmonic decay into two phonons with 
identical frequencies and opposite 
momenta~\cite{Klemens1966,Cardona1984}:      
\begin{equation}
\omega(T)=\omega_0-\omega_2[1+\frac{2}{e^{\hbar\omega_0/2k_BT}-1}],
\label{eq:energyTwo}
\end{equation}
and
\begin{equation}
\Gamma(T)=\Gamma_0+\Gamma_2[1+\frac{2}{e^{\hbar\omega_0/2k_BT}-1}].
\label{eq:gammaTwo}
\end{equation}

The fitting results are summarized in Table~\ref{tab:P}. Small 
deviations between the measured energy and the fitting curve for the 
B$_{1g}$ mode could be due to an additional decay channels, for example decay into one acoustic and one optical mode.

\begin{table}
\caption{\label{tab:P}The fitting parameters for the energy and FWHM of the four Raman-active optical phonon modes.}
\begin{ruledtabular}
\begin{tabular}{ccccc}
Mode&$\omega_0$&$\omega_2$&$\Gamma_0$&$\Gamma_2$\\
\hline
E$_{g}^{(1)}$&138.4$\pm$0.1&0.70$\pm$0.03&1.7$\pm$0.5&0.3$\pm$0.1\\
B$_{1g}$&167.92$\pm$0.01&0.212$\pm$0.002&6.08$\pm$0.02&0.094$\pm$0.005\\
A$_{1g}$&270.15$\pm$0.04&2.27$\pm$0.02&3.6$\pm$0.1&1.71$\pm$0.07\\
E$_{g}^{(2)}$&303.32$\pm$0.02&2.67$\pm$0.01&2.98$\pm$0.05&0.75$\pm$0.03\\
\end{tabular}
\end{ruledtabular}
\end{table}

The integrated intensity of the A$_{1g}$ and E$_{g}^{(2)}$ phonon modes has more than 50\% increase on cooling, in contrast to the behavior of the B$_{1g}$ phonon mode, whose integrated intensity is nearly temperature-independent [Fig.~\ref{fig:P_Para}(c)]. The increase of the integrated intensity on cooling suggests a coupling of a CF transition to these two phonon modes~\cite{Cooper1987}. This coupling is enhanced when the energies of the CF splitting and the phonon modes are close. Indeed, such a CF excitation, 239\,cm$^{-1}$ at 13\,K, exists. The mechanism of this coupling will be discussed in Subsection.~\ref{subsubsec:Coupling}.

Because the FQ order parameter is of B$_{1g}$ symmetry~\cite{Fisher2018}, the energy and lifetime of the B$_{1g}$ phonon mode are expected to exhibit anomalies across T$_Q$ due to electron-phonon coupling. Moreover, breaking of the four-fold rotational symmetry should split the two E$_{g}$ phonon modes~\cite{Zhang2016}. However, as shown in the insets of Fig.~\ref{fig:P_Para}, B$_{1g}$ and E$_{g}$ phonon modes do not exhibit significant anomaly across T$_Q$. E$_{g}$ phonon modes do not show notable splitting at 4.5\,K, either [Fig.~\ref{fig:P}]. The splitting of the E$_{g}^{(1)}$ phonon mode is challenging to observe due to its weak intensity. Because the FWHM of the E$_{g}^{(2)}$ phonon mode is 4\,cm$^{-1}$ at 4.5\,K, we set the upper limit of the splitting of the E$_{g}$ phonon modes to be about 4\,cm$^{-1}$ at 4.5\,K.

\subsection{Quadrupolar fluctuations\label{subsec:Q}}

In the tetragonal phase above $T_{Q}$, the four-fold rotational symmetry along the z-axis is preserved and the CF ground state supports no static xy-plane quadrupole moment. 
However, dynamical quadrupolar fluctuations with zero time average quadrupolar moment are allowed~\footnote{Detailed discussion of the symmetry 
breaking and the wavefunction mixing can be found in 
Subsection.\ref{subsec:CF}}.  

\begin{figure}
\includegraphics[width=0.44\textwidth]{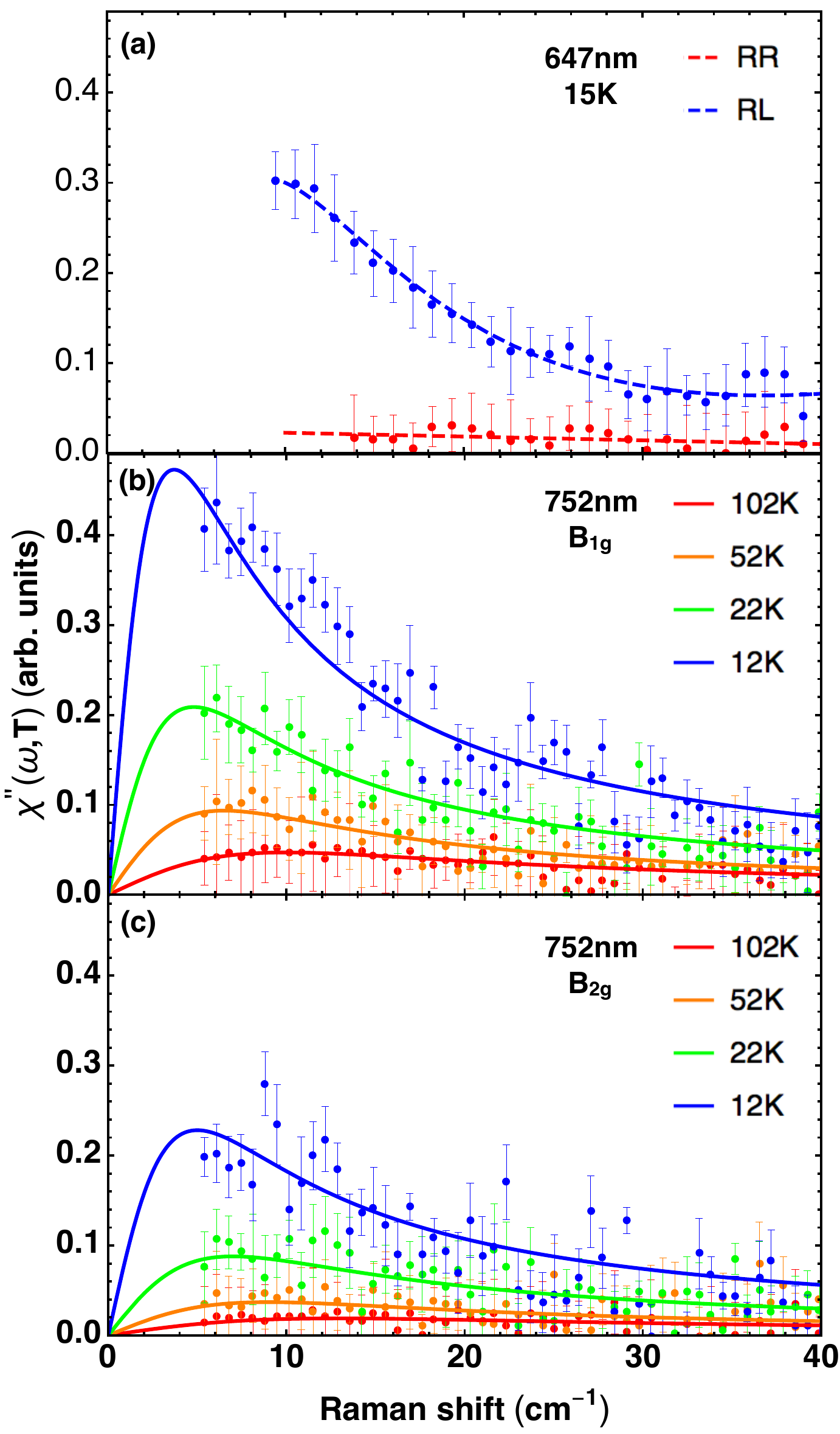}
\caption{\label{fig:CF1}
Raman response $\chi''(\omega,T)$ for (a) RR and RL scattering geometries with the 647.1\,nm excitation, (b) X'Y' geometry with the 752.5\,nm excitation, and (c) XY geometry with the 752.5\,nm excitation. The solid lines are Drude fits, Eq.~(\ref{eq:Drude}). The error bars represent one standard deviation.}
\end{figure}

In Fig.~\ref{fig:CF1} we show the spectra of low-energy quadrupolar fluctuations. 
They are present in RL geometry yet absent in RR geometry [Fig.~\ref{fig:CF1}(a)]. 
By group theory, the absence of A$_{1g}$ and A$_{2g}$ components indicates that the CF ground state is a quasi-quartet composed of one $\Gamma_6$ and one $\Gamma_7$ doublets.

The static Raman susceptibility $\chi_{\mu}(0,T)$ in the symmetry channel $\mu$ ($\mu$\,=\,B$_{1g}$ or B$_{2g}$) can be obtained from the Raman response $\chi_{\mu}^{\prime\prime}(\omega,T)$ by virtue of the Kramers-Kronig relation~\cite{Thorsmolle2016,Gallais2016}:
\begin{equation}
\chi_{\mu}(0,T)=\frac{2}{\pi}\int_{0}^{\omega_{max}} \frac{\chi_{\mu}^{\prime\prime}(\omega,T)}{\omega} d\omega~,
\label{eq:KK}
\end{equation}
in which we choose the upper cutoff for the spectra of fluctuations at $\omega_{max}$\,=\,40\,cm$^{-1}$, see Fig.~\ref{fig:CF1}.

We use Drude lineshape
\begin{equation}
\chi_{\mu}^{\prime\prime}(\omega,T)\propto\frac{Q_{\mu}^2\omega}{\omega^2+\gamma_{\mu}^2}
\label{eq:Drude}
\end{equation}
to extrapolate the Raman response below the instrumental cutoff 
5\,cm$^{-1}$. 
In Eq.~(\ref{eq:Drude}), $Q_{\mu}$ is the magnitude of 
the quadrupolar moment, and $\gamma_{\mu}$ reflects the decay rate. 
In the Raman scattering process light couples to the system's charge 
quadrupole moment.

\begin{figure}[b]
\includegraphics[width=0.48\textwidth]{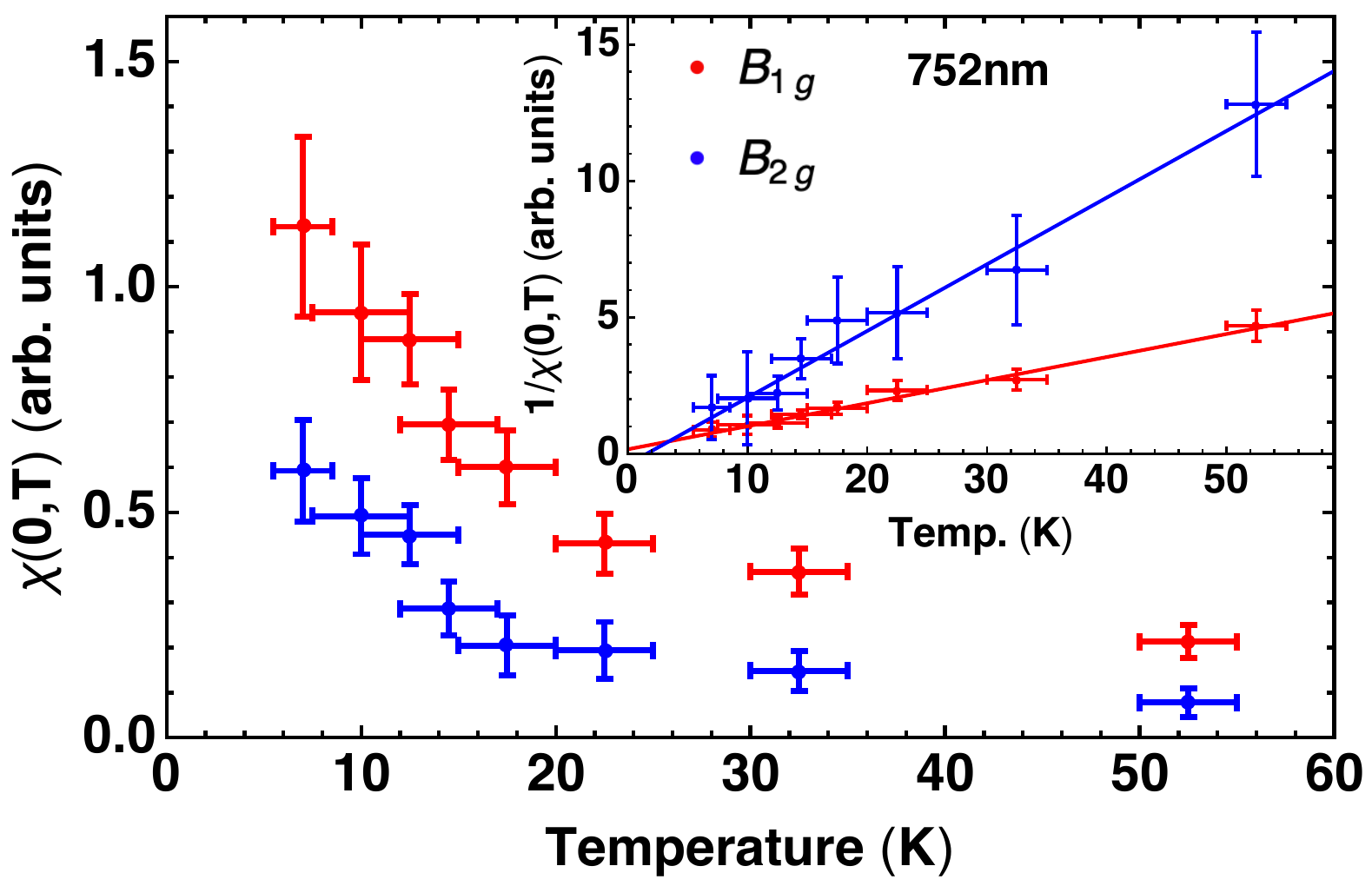}
\caption{\label{fig:CF1_Para}Temperature dependence of the static electronic Raman susceptibility $\chi(0,T)$ for B$_{1g}$ (red) and B$_{2g}$ (blue) quadrupole channels derived from Raman response shown in Fig.~\ref{fig:CF1}. Inset: temperature dependence of $1/\chi(0,T)$. The solid lines are Curie-Weiss fits, Eq.~(\ref{eq:SRS}). The vertical error bars represent one standard deviation; the horizontal ones are derived from the uncertainty of laser heating estimation.}
\end{figure}

Theoretically, the low-energy Raman response in the quadrupolar channels contains both the lattice and the electronic contributions~\cite{Thorsmolle2016,Gallais2016}. However, the energy of lattice fluctuations is much lower than the instrumental cutoff (5\,cm$^{-1}$), and Eq.~(\ref{eq:Drude}) only takes into account of the electronic contribution. Thus, only electronic quadrupole fluctuations are included in the derived susceptibility $\chi_{\mu}(0,T)$.

The obtained temperature dependence of the static electronic Raman susceptibilities for both B$_{1g}$ and B$_{2g}$ quadrupole channels are shown in Fig.~\ref{fig:CF1_Para}. The static Raman susceptibility $\chi_{\mu}(0,T)$ obeys Curie-Weiss temperature dependence
\begin{equation}
\chi_{\mu}(0,T)\propto\frac{Q_{\mu}^2}{T-T^{\mu}_{W}}~, 
\label{eq:SRS}
\end{equation}
where $T^{\mu}_{W}$ is the Weiss temperature:
\begin{equation}
T^{\mu}_{W}=\kappa_{\mu}Q_{\mu}^2~, 
\label{eq:Tw}
\end{equation}
in which $\kappa_{\mu}$ measures the strength of the electronic intersite quadrupolar interactions.

By fitting the data, the ratio of $Q_{B_{1g}}$ to $Q_{B_{2g}}$ is 
determined to be nearly 1.4. The derived Weiss temperatures, 
T$^{B_{1g}}_{W}$\,$\sim$\,$-2$ and 
T$^{B_{2g}}_{W}$\,$\sim$\,$+2$\,K
\footnote{We note that in 
the elastoresistivity studies, the Weiss temperature obtained from a 
fit to the static quadrupole-strain susceptibility is also a measure 
of an extrapolated bare electronic quadrupolar transition temperature 
because of clamped lattice~\cite{Fisher2018}. This is the reason why 
the Weiss temperatures extrapolated from the Raman susceptibility and 
from quadrupole-strain susceptibility agree.}. The nearly exact Curie 
law is not surprising because both direct-exchange and super-exchange 
between local quadrupolar moments are expected to be vanishingly weak 
due to compactness of the $f$-orbitals. Although itinerant electrons 
provide effective coupling between the local magnetic dipole moments 
at Yb$^{3+}$ sites, and the resulting RKKY interaction~\footnote{See 
Ref.~\cite{RKKY} and the references therein.} leads to AFM order 
below T$_{N1}$=6.5\,K, these itinerant electrons do not provide a 
significant effective coupling between the local electric quadrupole moments at 
Yb$^{3+}$ sites.                     

The true B$_{1g}$-symmetry FQ order develops at 
T$_{Q}$\,=\,10\,K~\cite{Fisher2018}, about 10\,K above the Weiss 
temperature T$^{B_{1g}}_{W}$. Because YRu$_2$Ge$_2$, the non-magnetic 
analog of the same structure, has no orthorhombic 
transition~\cite{Bouvier1996,Fisher2018}, the quadrupolar 
fluctuations of YbRu$_2$Ge$_2$ lattice themselves should have 
little tendency towards a structural instability. 
Nevertheless, coupling between the lattice strain fields and the local electronic quadrupole moments of the same symmetry enhances the transition temperature~\cite{Bohmer2015,Bohmer2016,Thorsmolle2016,Gallais2016}:
\begin{equation}
T^{\mu}_{Q} = T^{\mu}_{W}+(\lambda_{\mu}^2/C_{\mu})Q_{\mu}^2 = (\kappa_{\mu}+\lambda_{\mu}^2/C_{\mu})Q_{\mu}^2~,
\label{eq:Diff}
\end{equation}
where $\lambda_{\mu}$ measures the coupling between the local 
charge quadrupole moments on Yb$^{3+}$ sites and the lattice strain 
fields, and $C_{\mu}$ is the symmetrized elastic constant, which is 
(C$_{11}$-C$_{12}$)/2 for the B$_{1g}$ channel or C$_{66}$ for the 
B$_{2g}$ channel~\cite{Mitsumoto2013}. The true quadrupolar 
transition temperature T$_Q$ equals to the largest of two T$^{\mu}_{Q}$. 
Because the FQ order in YbRu$_2$Ge$_2$ has B$_{1g}$ symmetry, the 
T$^{B_{1g}}_{Q}$\,=\,T$_Q$ and non-realized T$^{B_{2g}}_{Q}$\,\textless\,T$_Q$.

Tuning an additional parameter (magnetic field, pressure or doping, for instance) may affect the electron-lattice coupling and induce a transition from B$_{1g}$ FQ ordering to B$_{2g}$ FQ ordering. Indeed, although T$_Q$ stays constant up to application 9\,GPa pressure with zero magnetic field~\cite{Nicklas2010} and increases with in-plane magnetic field at ambient pressure~\cite{Jeevan2006}, experimental results do show suppression of T$_Q$ by Si~\cite{Prasad2010} or Rh~\cite{ZXShen2019} doping, and by applying magnetic field under 1.23\,GPa pressure~\cite{Nicklas2010}. These results suggest a competition between B$_{1g}$- and B$_{2g}$-symmetry FQ order.

\subsection{Crystal-Field Excitations\label{subsec:CF}}

Within the $^2F_{7/2}$ multiplet, there are three CF excitations 
corresponding to transitions from the CF ground state to the three CF 
excited states. 
From group theoretical considerations~\cite{Koster1963}, the CF transitions between levels of the same symmetry (i.e. $\Gamma_6\rightarrow\Gamma_6$ or $\Gamma_7\rightarrow\Gamma_7$) contain A$_{1g}$, A$_{2g}$ and E$_{g}$ representations, whereas those between levels of different symmetry (i.e. $\Gamma_6\rightarrow\Gamma_7$ or $\Gamma_7\rightarrow\Gamma_6$) contain B$_{1g}$, B$_{2g}$ and E$_{g}$ symmetry representations. The Raman intensities in different symmetry channels may vary due to matrix element effect.

\begin{figure}
\includegraphics[width=0.44\textwidth]{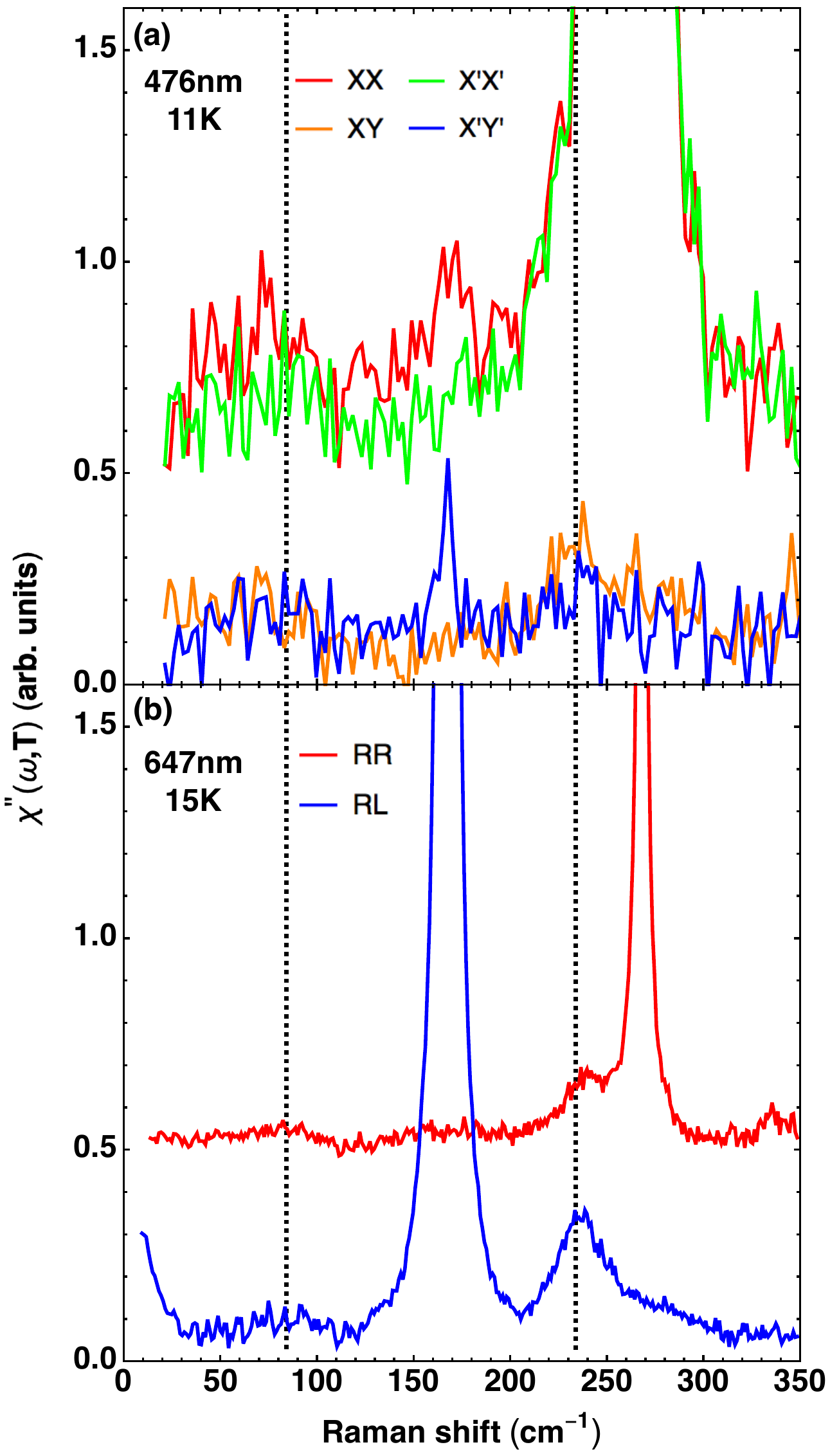}
\caption{\label{fig:CF23}Raman response $\chi''(\omega)$ of the CF excitations from the quasi-quartet to the remaining two CF levels at low temperature. The dashed lines indicate the position of the two CF transitions. (a) The spectra for four linear polarizations measured at 11\,K with the 476.2\,nm excitation. The XX and X'X' spectra are offset by 0.5 arbitrary units (arb. units). The spectral resolution is 3.5\,cm$^{-1}$. (b) The spectra for two circular polarizations measured at 15\,K with the 647.1\,nm excitation. The RR spectrum is offset by 0.5 arbitrary unit. The spectral resolution is 2.5\,cm$^{-1}$.}
\vspace{-2mm}
\end{figure}

The lowest-energy CF transition, namely the transition between the two quasi-degenerate Kramers doublet does not clearly exhibit itself in the low-energy Raman spectra [Fig.~\ref{fig:CF1}]. 
The CF excitations from the ground state to the remaining two higher 
energy states are shown in Fig.~\ref{fig:CF23} at 95\,cm$^{-1}$ and 
239\,cm$^{-1}$. 
These two transitions are expected to appear in all Raman-active symmetry channels, 
because the two low-lying doublets within the quasi-quartet have 
roughly the same population at 11\,K.  
With the 476.2\,nm excitation, the 95\,cm$^{-1}$ transition indeed appears as a weak peak for four linear polarizations, while the 239\,cm$^{-1}$ transition overlaps with the strong A$_{1g}$ phonon mode. With the 647.1\,nm excitation, instead, the 95\,cm$^{-1}$ transition becomes weaker, but the 239\,cm$^{-1}$ transition is identifiable, manifesting itself as a peak in the RL spectrum and a shoulder in the RR spectrum.

The symmetry of the CF levels are assigned by the following argument: because YbRh$_2$Si$_2$ and YbIr$_2$Si$_2$, which have the same lattice structure as YbRu$_2$Ge$_2$, both have a $\Gamma_6$ CF ground state~\cite{Kutuzov2011,Leushin2008}, we suggest that the CF ground state of YbRu$_2$Ge$_2$ is also of $\Gamma_6$ symmetry (denoted as $\Gamma_6^{(1)}$). The other Kramers doublet within the quasi-quartet, in turn, is of $\Gamma_7$ symmetry (denoted as $\Gamma_7^{(1)}$).

The small difference of the excitation energy measured in RR and RL 
geometries near 239\,cm$^{-1}$ [Fig.~\ref{fig:CF23}(b)] serves as an 
estimation of the quasi-quartet splitting. Using the Lorentzian fits, 
we find that the excitation energy measured in RR geometry is higher 
by 2\,$\pm$\,1\,cm$^{-1}$ than that in RL geometry at 25\,K and 
15\,K. Therefore, the symmetry of the CF state at 239\,cm$^{-1}$ is 
defined to be $\Gamma_6$ (denoted as $\Gamma_6^{(2)}$), and the quasi-quartet splitting is estimated to be 2\,$\pm$\,1\,cm$^{-1}$. Because there are only two $\Gamma_6$ and two $\Gamma_7$ states within the $^2F_{7/2}$ multiplet, the CF state at 95\,cm$^{-1}$ can only be of $\Gamma_7$ symmetry (denoted as $\Gamma_7^{(2)}$).

The energies of the CF excitations at 15\,K are summarized in 
Table~\ref{table:ModeSummary} 
\footnote{We note that would the CF ground state be $\Gamma_7$ 
symmetry, the symmetry of the CF states at 2\,cm$^{-1}$, 
95\,cm$^{-1}$ and 239\,cm$^{-1}$ should instead be $\Gamma_6$, 
$\Gamma_6$ and $\Gamma_7$, respectively.}.    

In an inelastic neutron scattering study of YbRu$_2$Ge$_2$, excitations at 89\,cm$^{-1}$, 170\,cm$^{-1}$ and 260\,cm$^{-1}$ are resolved at 5\,K with the magnitude of momentum transfer being $\sim$1.9\AA$^{-1}$ (Ref.~\cite{Jeevan2010}). 
Their data well match our assignments; the comparison is shown in Table~\ref{table:ModeSummary}. 
This consistency not only supports our assignments, but also suggests that the CF excitations and optical phonon modes have little dispersion.

\begin{figure}[t]
\includegraphics[width=0.48\textwidth]{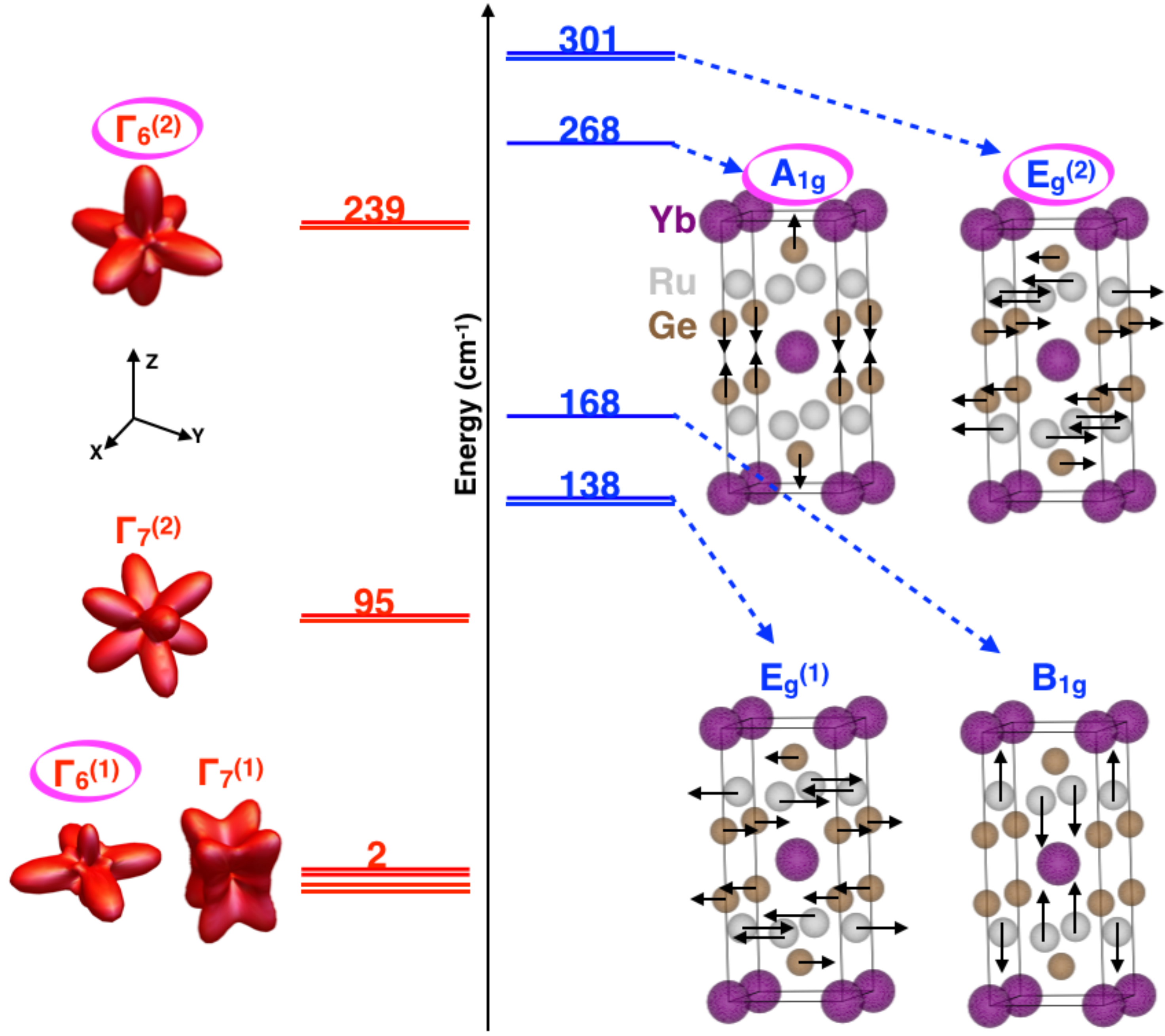}
\caption{\label{fig:EP} Schematic energy diagram of the CF states (red horizontal lines) and the phonon modes (blue horizontal lines). The coupled CF transition and phonon modes are circled in purple. On the left are the angular electron-cloud distributions of the four CF states; on the right are the vibration patterns of the phonon modes.}
\end{figure}

\subsubsection{Angular Electron-Cloud Distribution of the Crystal-Field States\label{subsubsec:Model}}

We use the following single-ion Hamiltonian to calculate the angular electron-cloud distribution at Yb sites:
\begin{equation}
H=H_{CF}+H_{B_{1g}}~.
\label{eq:H}
\end{equation}

The first term in Eq.~(\ref{eq:H})
\begin{equation}
H_{CF}=B_2^0\hat{O}_2^0+B_4^0\hat{O}_4^0+B_4^4\hat{O}_4^4+B_6^0\hat{O}_6^0+B_6^4\hat{O}_6^4
\label{eq:HCF}
\end{equation}
is the general expression for a CF potential of tetragonal site symmetry~\cite{Fischer1987}. The $\hat{O}^0_2$, $\hat{O}^0_4$, $\hat{O}^4_4$, $\hat{O}^0_6$, and $\hat{O}^4_6$ operators are Stevens operators~\cite{Stevens1952}. The five $B$'s are the CF coefficients.

From the CF level diagram, we cannot uniquely determine the CF 
Hamiltonian and wavefunctions if we assume tetragonal site symmetry. 
Hence, we approximate the real tetragonal CF potential with a dominating cubic CF potential~\cite{Lea1962} plus a small $\hat{O}^0_2$ axial term:
\begin{equation}
H_{Tetra}=B_2^0\hat{O}_2^0+B_4(\hat{O}^0_4+5\hat{O}^4_4)+B_6(\hat{O}^0_6-21\hat{O}^4_6)~.
\label{eq:HT}
\end{equation}

A cubic CF potential would split the $^2F_{7/2}$ multiplet into one 
quartet $\Gamma_8$, one doublet $\Gamma_7$, and one doublet 
$\Gamma_6$ states of O$_{h}$ group. Reducing the cubic symmetry to 
the tetragonal symmetry, the quartet $\Gamma_8$ state of O$_{h}$ 
group would be split into one $\Gamma_7$ and one $\Gamma_6$ states of 
D$_{4h}$ group. Because YbRu$_2$Ge$_2$ has a quasi-quartet CF ground 
state, it is possible that this quasi-quartet is induced by a small 
tetragonal perturbation to a large cubic CF potential. This small 
perturbation is represented by the first term in Eq.~(\ref{eq:HT}). 
We cannot rule out an alternative scenario that the quasi-quartet CF ground state of YbRu$_2$Ge$_2$ is of accidental degeneracy, rather than derived from the quartet $\Gamma_8$ state of cubic symmetry. 
Nevertheless, the Hamiltonian $H_{Tetra}$ preserves the 4-fold 
rotational symmetry along z-axis, and is sufficient to provide 
qualitative insights.  
In Appendix~\ref{sec:AB2g} we show that based on our assumption, the 
ratio of $Q_{B_{1g}}$ to $Q_{B_{2g}}$ is calculated to be 1.34, close 
to the experimentally determined ratio of 1.4 
[Subsection.\ref{subsec:Q}]. This consistency supports our choice of 
Eq.~(\ref{eq:HT}). 
Experimentally, the wavefunction of the CF ground 
state could be determined by core-level non-resonant inelastic X-ray 
scattering, which has been used for Ce-based heavy 
fermion systems~\cite{Sundermann2019}.

The second term in Eq.~(\ref{eq:H})
\begin{equation}
H_{B_{1g}}=\frac{V}{2}(\hat{J}_x^2-\hat{J}_y^2)=\frac{V}{2}(\hat{J}_+^2+\hat{J}_-^2)
\label{eq:HB1g}
\end{equation}
represents the effective quadrupole-field (QF) potential of B$_{1g}$ symmetry. $V$ measures the strength of the QF potential.

Above T$_Q$, there is no static B$_{1g}$ QF potential and we define $H=H_{Tetra}$. 
We diagonalize $H_{Tetra}$ in the basis of $|J,m_J\rangle$, where $J=7/2$ and $m_J$ are the quantum numbers of $\hat{\mathbf{J}}$ and $\hat{J}_z$, respectively. After diagonalization, the CF transition energies can be expressed in terms of $B_2^0$, $B_4$, and $B_6$. We fit the experimentally determined CF level diagram by these three adjustable parameters. There are four sets of parameters which reproduce the level diagram, and we choose the set with the smallest $B_2^0$ value. The fitting results thus are $B_2^0$\,=\,-0.164\,cm$^{-1}$, $B_4$=0.0518\,cm$^{-1}$, and $B_6$=-0.00442\,cm$^{-1}$. 
The corresponding angular electron-cloud distribution of the CF states is plotted in Fig.~\ref{fig:EP}.

\begin{figure}
\includegraphics[width=0.48\textwidth]{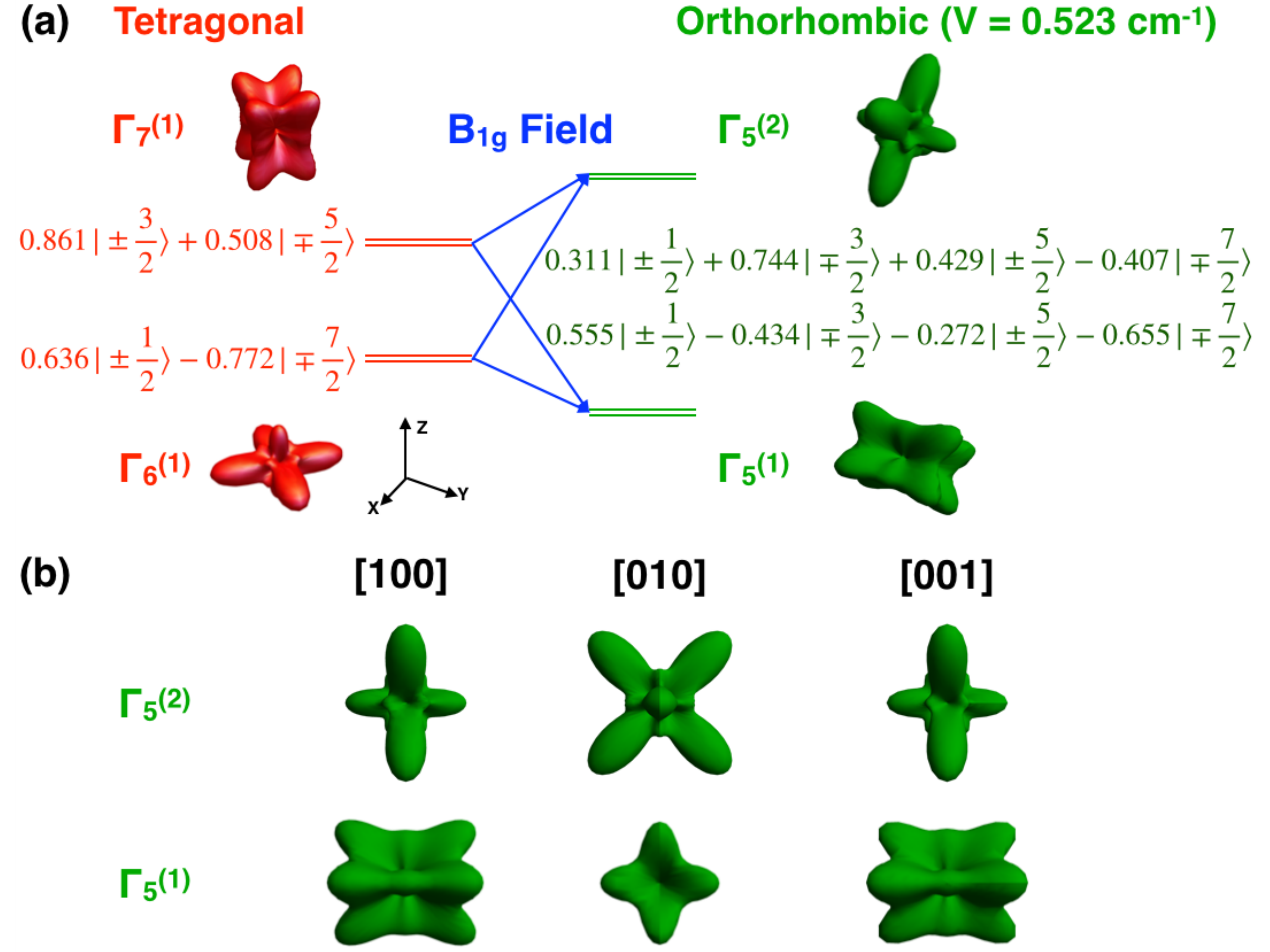}
\caption{\label{fig:FQ} (a) The effect of the B$_{1g}$ quadrupole-field potential on the ground quasi-quartet. The $\Gamma_6^{(1)}$ and $\Gamma_7^{(1)}$ doublets of the D$_{4h}$ group are mixed to form the $\Gamma_5^{(1)}$ and $\Gamma_5^{(2)}$ doublets of the D$_{2h}$ group. The wavefunctions are expressed in the basis of $|J=7/2,m_J\rangle$. (b) The angular electron-cloud distribution of the $\Gamma_5^{(1)}$ and $\Gamma_5^{(2)}$ doublets viewed from three orthogonal directions.}
\end{figure}

Below T$_Q$, there is a finite static B$_{1g}$ QF potential, here we 
define $H=H_{Tetra}+H_{B_{1g}}$. 
We assume that the values of $B_2^0$, $B_4$, and $B_6$ do not change. 
We diagonalize $H$ in the basis of $|J,m_J\rangle$, and after 
diagonalization, the CF transition energies can be expressed in terms 
of $V$. We find that $V$=0.523\,cm$^{-1}$ renders an additional 
2\,cm$^{-1}$ splitting of the ground quartet. In Fig.~\ref{fig:FQ}, 
we plot the angular electron-cloud distribution of the ground quartet 
for $V$=0.523\,cm$^{-1}$. The charge distribution looks different 
from [100] and  [010] directions because the $\Gamma_5^{(1)}$ and 
$\Gamma_5^{(2)}$ doublets carry B$_{1g}$ quadrupole moment. 
Furthermore, the quadrupole moment carried by $\Gamma_5^{(1)}$ state 
and that carried by $\Gamma_5^{(2)}$ state have approximately same 
magnitude but an opposite sign~\footnote{Strictly speaking, the 
quadrupole moments have the same magnitude only if we ignore the 
contribution by higher energy CF states.}.   

The FQ phase transition reflects the competition between the entropy and energy terms in the Helmholtz free energy of the system. Above T$_Q$, the entropy term dominates and the system prefers a quasi-degenerate CF ground state. Below T$_Q$, instead, the system pursues lowest possible energy, and an orthorhombic quadrupolar field fulfills the goal: this field mixes the wavefunctions of the quasi-degenerate $\Gamma_6^{(1)}$ and $\Gamma_7^{(1)}$ states, increasing their separation and in turn reducing the ground state energy. In view of group-theoretical considerations, the $\Gamma_6$ and $\Gamma_7$ irreducible representations of the D$_{4h}$ group become the $\Gamma_5$ representation of the D$_{2h}$ group. Correspondingly, the $\Gamma_6^{(1)}$ and $\Gamma_7^{(1)}$ states of the D$_{4h}$ tetragonal phase are mixed by the Hamiltonian $H_{B_{1g}}$, and become the $\Gamma_5^{(1)}$ and $\Gamma_5^{(2)}$ states of the D$_{2h}$ orthorhombic phase.

There are two obvious choices for the macroscopic order parameter of 
the B$_{1g}$-symmetry FQ phase. 
First is the quadrupole moment per unit volume:
\begin{equation}
\Psi\propto(n_{\Gamma_5^{(1)}}-n_{\Gamma_5^{(2)}})Q_{B_{1g}}~,
\label{eq:MOP1}
\end{equation}
where $n_{\Gamma_5^{(1)}}$ and $n_{\Gamma_5^{(2)}}$ are the occupancy for the $\Gamma_5^{(1)}$ and $\Gamma_5^{(2)}$ states, respectively.
The second choice is the lattice orthorhombicity, which 
couples to the quadrupolar order:
\begin{equation}
\Psi\propto\frac{a-b}{a+b}~,
\label{eq:MOP2}
\end{equation}
where $a$ and $b$ are the in-plane lattice constant. 
The orthorhombicity as a function of temperature has been measured by 
X-ray diffraction~\cite{Fisher2018}.  

\subsubsection{Coupling between the Crystal-Field Transition and the Phonon Modes\label{subsubsec:Coupling}}

The coupling between the $\Gamma_6^{(1)}\rightarrow\Gamma_6^{(2)}$ CF transition and the A$_{1g}$ and E$_{g}^{(2)}$ phonon modes originates from the modulation of the electron-cloud distribution of CF states by lattice vibration [Fig.~\ref{fig:EP}]. Such coupling is allowed by group theory because $\Gamma_6\otimes\Gamma_6\,=\,A_{1g}\oplus A_{2g}\oplus E_{g}$. We note that the phonon energy and linewidth can be well accounted for by the anharmonic decay model [Fig.~\ref{fig:P_Para}(a) and (b)], suggesting that renormalization due to electron-phonon coupling is small. In the Appendix~\ref{sec:ADerivation} we show that for small coupling strength, the temperature dependence of the integrated intensity of the phonon modes, $I.I.(T)$, has the following phenomenological expression:
\begin{equation}
I.I.(T)=Af_{(1)}(T)[1-f_{(2)}(T)]+B~,
\label{eq:Int}
\end{equation}
where $A$ and $B$ are two constants; $f_{(1)}(T)=2/Z(T)$ measures the occupancy of the $\Gamma_6^{(1)}$ CF state, and $f_{(2)}(T)=2e^{-E_4/k_BT}/Z(T)$ measures the occupancy of the $\Gamma_6^{(2)}$ CF state. $Z=2\sum_{i=1}^{4}e^{-E_i/k_BT}$ is the partition function; $E_1$=0\,cm$^{-1}$, $E_2$=2\,cm$^{-1}$, $E_3$=95\,cm$^{-1}$ and $E_4$=239\,cm$^{-1}$ are the energies of the CF levels [Table~\ref{table:ModeSummary}].

In Eq.~(\ref{eq:Int}), the constant $B$ represents the temperature-independent spectral weight of the phonon mode. Without the interaction $v$ and in the absence of a phase transition, the integrated intensity of the phonon modes is expected to be temperature-independent. The first term, which is temperature-dependent, can be interpreted as the spectral weight transferred from the CF mode to the phonon mode. This transferred spectral weight is proportional to the occupancy of the ground CF state $\Gamma_6^{(1)}$, and the un-occupancy of the excited CF state $\Gamma_6^{(2)}$. The constant $A$ is a measure of the transferred spectral weigh at zero-temperature~\footnote{We note that there is no general sum rule for Raman 
spectroscopy.}.

Because of the phase transition at T$_Q$=10\,K, Eq.~(\ref{eq:Int}) is only valid above 10\,K. In addition, group theory allows the $\Gamma_7^{(1)}\rightarrow\Gamma_6^{(2)}$ CF mode to couple to the E$_{g}^{(2)}$ phonon mode, which is not considered by simplified Eq.~(\ref{eq:Int}). Because the splitting between the $\Gamma_6^{(1)}$ and $\Gamma_7^{(1)}$ states is only 2\,cm$^{-1}$, including the contribution from the $\Gamma_7^{(1)}\rightarrow\Gamma_6^{(2)}$ CF mode will only influences the fitting curve at temperature much lower than T$_Q$, a temperature range where Eq.~(\ref{eq:Int}) is invalid.

We use Eq.~(\ref{eq:Int}) to fit the phonon intensity data above 10\,K in Fig.~\ref{fig:P_Para}(c). For the A$_{1g}$ phonon mode, $A=3.14\pm0.08$ and $B=0.06\pm0.03$; for the E$_{g}^{(2)}$ phonon mode, $A=2.08\pm0.05$ and $B=0.35\pm0.02$. These values show that at low-temperature, the integrated intensity of the A$_{1g}$ and E$_{g}^{(2)}$ modes is mainly contributed by the transferred spectral weight. The fitting curves match the data well, which further supports our CF level scheme.

\section{Conclusion\label{sec:Conclusion}}
In summary, the Raman scattering study of YbRu$_2$Ge$_2$ focuses on the origin of the ferroquadrupolar transition, as well as on the spectroscopy of phonons and CF excitations within the $^2F_{7/2}$ ground multiplet of Yb$^{3+}$ ion.

The deduced CF level scheme verifies the proposed quasi-quartet ground state, and we estimate that the splitting between two quasi-degenerate Kramers doublets is about 2\,cm$^{-1}$. The static electronic Raman susceptibilities in both B$_{1g}$ and B$_{2g}$ quadrupole channels essentially exhibit Curie law, signifying relatively strong coupling to the lattice in the B$_{1g}$-symmetry channel that enhances the vanishingly small electronic Weiss temperature to the temperature of quadrupole phase transition at 10\,K. 

The temperature dependence of the energy and FWHM of the observed phonon modes are described by anharmonic decay model. The integrated intensities of the A$_{1g}$ and E$_{g}^{(2)}$ phonon modes show more than 50\% enhancement on cooling, which is explained by strong coupling between these phonons and the CF transitions with similar energies. 

\begin{acknowledgments}
We thank K. Haule and G. Khanal for discussions. The spectroscopic work at Rutgers (M.Y. and G.B.) was supported by NSF Grant No. DMR-1709161. The work at Stanford (E.W.R. and I.R.F.), including the crystal growth and characterization, was supported by the Gordon and Betty Moore Foundation Emergent Phenomena in Quantum Systems Initiative through Grant No. GBMF4414. G.B. also acknowledges the QuantEmX grant from ICAM and the Gordon and Betty Moore Foundation through Grant No. GBMF5305 allowing to make G.B. a collaborative visit to Stanford. Work at NICBP was supported by IUT23-3 grant.     
\end{acknowledgments}

\appendix

\section{The effect of the B$_{2g}$ quadrupole-field potential\label{sec:AB2g}}

For completeness, we analyze here the effect of the B$_{2g}$ QF potential on the ground quasi-quartet. Following the treatment in Subsection.~(\ref{subsubsec:Model}), we take $H=H_{Tetra}+H_{B_{2g}}$, where~\footnote{Asterisks are used to distinguish the symbols used for $H_{B_{2g}}$ from those used for $H_{B_{1g}}$.}
\begin{equation}
H_{B_{2g}}=\frac{V^*}{2}(\hat{J}_x\hat{J}_y+\hat{J}_y\hat{J}_x)=\frac{V^*}{4i}(\hat{J}_+^2-\hat{J}_-^2)~.
\label{eq:HB2g}
\end{equation}

We find that for B$_{2g}$ potential, $V^*$\,=\,0.668\,cm$^{-1}$ renders a 2\,cm$^{-1}$ additional splitting of the ground quartet. On the contrary, for B$_{1g}$ potential $V$=0.523\,cm$^{-1}$ renders a 2\,cm$^{-1}$ additional splitting of the ground quartet. Hence smaller B$_{1g}$ QF potential is needed to induce the same additional splitting of the quasi-quartet. This result is consistent with the conclusion that the coupling between the local quadrupole moments and the lattice strain field is stronger in the B$_{1g}$ channel than in the B$_{2g}$ channel.

\begin{figure}
\includegraphics[width=0.48\textwidth]{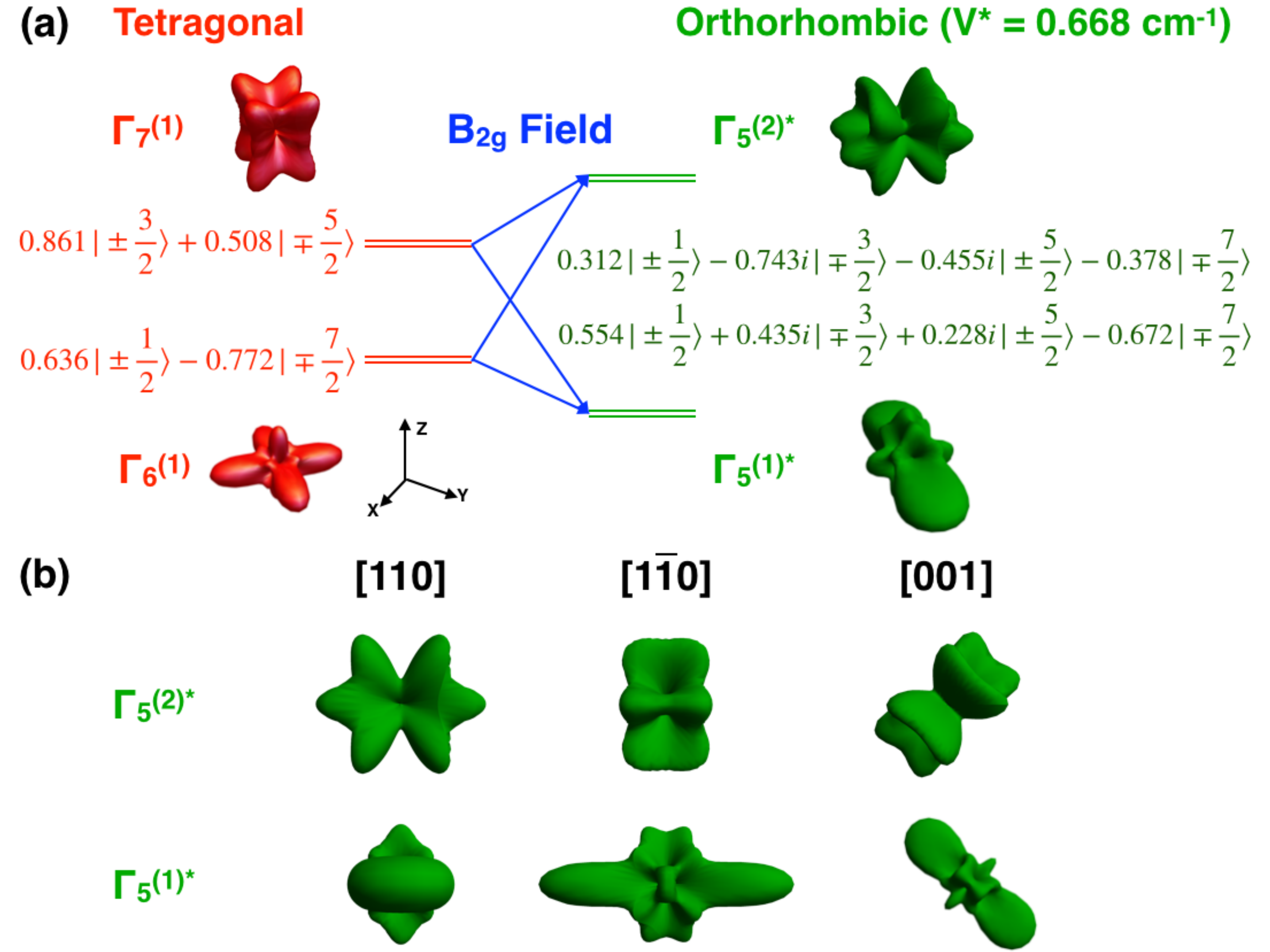}
\caption{\label{fig:FQB2g} (a) The effect of the B$_{2g}$ quadrupole-field potential on the ground quasi-quartet. The $\Gamma_6^{(1)}$ and $\Gamma_7^{(1)}$ doublets of the D$_{4h}$ group are mixed to form the $\Gamma_5^{(1)*}$ and $\Gamma_5^{(2)*}$ doublets of the D$_{2h}$ group. The wavefunctions are expressed in the basis of $|J=7/2,m_J\rangle$. (b) The angular electron-cloud distribution of the $\Gamma_5^{(1)*}$ and $\Gamma_5^{(2)*}$ doublets viewed from three orthogonal directions.}
\end{figure}

In Fig.~\ref{fig:FQB2g}, we plot the angular electron-cloud distribution of the ground quartet for $V^*$\,=\,0.668\,cm$^{-1}$. The charge distribution looks different from [110] and  [1$\overline{1}$0] directions because the $\Gamma_5^{(1)*}$ and $\Gamma_5^{(2)*}$ doublets carry B$_{2g}$ quadrupole moment.

The traceless tensor of the electric quadrupole moments~\cite{Jackson1999Quadrupole}, written in Cartesian coordinate with arbitrary units, for the $\Gamma_5^{(1)}$ wavefunction generated by the B$_{1g}$ field, Eq.~(\ref{eq:HB1g}), has the following values:
\begin{eqnarray}
\begin{pmatrix}
	0.655&0&0\\
	0&-0.346&0\\
	0&0&-0.309\\
\end{pmatrix}
\end{eqnarray}

Hence the magnitude of the B$_{1g}$-symmetry electric quadrupole moment $Q_{B_{1g}}$ of the charge distribution of the $\Gamma_5^{(1)}$ wavefunction has a value of 1.00 when $V$=0.523\,cm$^{-1}$. 

The same tensor for the $\Gamma_5^{(1)*}$ wavefunction generated by the B$_{2g}$ field, Eq.~(\ref{eq:HB2g}) has the following values:
\begin{eqnarray}
\begin{pmatrix}
	0.176&-0.748&0\\
	-0.748&0.176&0\\
	0&0&-0.352\\
\end{pmatrix}
\end{eqnarray}

The magnitude of the B$_{2g}$-symmetry electric quadrupole moment $Q_{B_{2g}}$ of the charge distribution of the $\Gamma_5^{(1)*}$ wavefunction is 0.748 when $V^*$\,=\,0.668\,cm$^{-1}$.

Therefore, for the same additional splitting of the ground quasi-quartet, the calculated ratio of $Q_{B_{1g}}$ to $Q_{B_{2g}}$ is 1.34. We recall that the experimentally determined ratio of $Q_{B_{1g}}$ to $Q_{B_{2g}}$ is 1.4. This consistency supports the assumptions made in Eq.~(\ref{eq:HT}), and shows that the wavefunctions we use are close to the real wavefunctions.

\section{Derivation of Eq.~(\ref{eq:Int})\label{sec:ADerivation}}

The Hamiltonian of the coupled CF transition and phonon mode can be written in second-quantization form as
\begin{equation}
H=\omega_{1}\hat{a}_1^{\dag}\hat{a}_1+\omega_{2}\hat{a}_2^{\dag}\hat{a}_2+\omega_{p}\hat{b}^{\dag}\hat{b}+v(\hat{a}_2^{\dag}\hat{a}_1-\hat{a}_1^{\dag}\hat{a}_2)(\hat{b}^{\dag}+\hat{b})~,
\label{eq:aH}
\end{equation}
where $\hat{a}^{\dag}$ and $\hat{a}$ are fermionic creation and destruction operators; $\hat{b}^{\dag}$ and $\hat{b}$ are bosonic creation and destruction operators. The first and second terms describe respectively the energy of the lower and upper CF level; the third term is the phonon energy; and the last term is the coupling between the CF transition and phonon mode. Coefficient $v$ measures the strength of the coupling, which we take as a real number.

The CF transition corresponds to a bubble-shape Feynman diagram of electron-hole pair. Neglecting self-energy, the propagator has the following form:
\begin{equation}
P(\omega,T)=\frac{f_1(T)[1-f_2(T)]}{\omega-\omega_{e}+i\epsilon}-\frac{[1-f_1(T)]f_2(T)}{\omega+\omega_{e}+i\epsilon})~,
\label{eq:aEP}
\end{equation}
where $f_1(T)$ and $f_2(T)$ are respectively the temperature-dependent occupancy of the lower and upper CF level; $\omega_{e}=\omega_{2}-\omega_{1}$ is the energy of the CF transition; and $\epsilon$ is an infinitesimal positive value.

The phonon propagator is
\begin{equation}
D(\omega,T)=\frac{1+n(\omega_{p},T)}{\omega-\omega_{p}+i\epsilon}-\frac{n(\omega_{p},T)}{\omega+\omega_{p}+i\epsilon}~,
\label{eq:aPP}
\end{equation}
where $n(\omega_{p},T)$ is the Bose distribution function.

The experimentally-measured scattering rate, $I(\omega,T)$, has the form
\begin{equation}
I(\omega,T)\sim\frac{1}{\pi}\Im T^{\dag}G(\omega,T)T~,
\label{eq:aRate1}
\end{equation}
where $T^{\dag}=\left(\begin{array}{cc}T_{p} &T_{e}\end{array}\right)$ is the vertex of the light scattering, and $G(\omega,T)$ is the Green's function of the Hamiltonian in Eq.~(\ref{eq:aH}). $G$ can be obtained by solving Dyson equation:
\begin{multline}
G(\omega,T)=\frac{1}{1-P(\omega,T)vD(\omega,T)v}\\
\begin{pmatrix}D(\omega,T) & D(\omega,T)vP(\omega,T)\\ P(\omega,T)vD(\omega,T) & P(\omega,T) \end{pmatrix}~.
\label{eq:aG}
\end{multline}
In the following derivation, we assume $v$ is small so that the prefactor in the above expression can be replaced by unity. Then the approximated form of $I(\omega,T)$ is
\begin{equation}
\frac{1}{\pi}\Im [T_{p}^2D(\omega,T)+2T_{p}T_{e}D(\omega,T)vP(\omega,T)+T_{e}^2P(\omega,T)]~.
\label{eq:aRate2}
\end{equation}

By virtue of Sokhotsky's formula
\begin{equation}
\lim_{\epsilon\to 0^+} \frac{1}{\omega\pm\omega_{0}+i\epsilon} = \textrm{p.v.}\frac{1}{\omega\pm\omega_{0}} - i\pi\delta(\omega\pm\omega_{0})~,
\label{eq:aSokhotsky}
\end{equation}
where p.v. stands for principle value, we derive from Eq.~(\ref{eq:aRate2}) the Stokes part of the scattering rate:
\begin{multline}
I(\omega,T)\sim T_{p}^2[1+n(\omega_{p},T)]\delta(\omega-\omega_{p})\\
+T_{e}^2f_1(T)[1-f_2(T)]\delta(\omega-\omega_{e})\\
+2T_{p}T_{e}v[1+n(\omega_{p},T)]f_1(T)[1-f_2(T)]\\
[\frac{\delta(\omega-\omega_{p})}{\omega_{p}-\omega_{e}}+\frac{\delta(\omega-\omega_{e})}{\omega_{e}-\omega_{p}}]~.
\label{eq:aRate3}
\end{multline}

Therefore, the phonon scattering rate, $I_p(\omega,T)$, is 
\begin{multline}
I_p(\omega,T)\sim T_{p}^2[1+n(\omega_{p},T)]\delta(\omega-\omega_{p})\\
+2T_{p}T_{e}v[1+n(\omega_{p},T)]f_1(T)[1-f_2(T)]\frac{\delta(\omega-\omega_{p})}{\omega_{p}-\omega_{e}}~,
\label{eq:aRateP1}
\end{multline}
which can be arranged into
\begin{equation}
T_{p}^2[1+n(\omega_{p},T)]\{1+2\frac{T_{e}}{T_{p}}\frac{v}{\omega_{p}-\omega_{e}}f_1(T)[1-f_2(T)]\}\delta(\omega-\omega_{p})~.
\label{eq:aRateP2}
\end{equation}

The phonon response function, $\chi''_p(\omega,T)$, in turn, is
\begin{multline}
\chi''_p(\omega,T)\sim\\
T_{p}^2\{1+2\frac{T_{e}}{T_{p}}\frac{v}{\omega_{p}-\omega_{e}}f_1(T)[1-f_2(T)]\}\delta(\omega-\omega_{p})~,
\label{eq:aResP}
\end{multline}
and integration of $\chi''_p(\omega,T)$ yields the integrated intensity of the phonon mode $I.I.(T)$:
\begin{equation}
I.I.(T)\sim T_{p}^2\{1+2\frac{T_{e}}{T_{p}}\frac{v}{\omega_{p}-\omega_{e}}f_1(T)[1-f_2(T)]\}~.
\label{eq:aInt1}
\end{equation}

Eq.~(\ref{eq:aInt1}) can be cast in a phenomenological form:
\begin{equation}
I.I.(T)=Af_1(T)[1-f_2(T)]+B~,
\label{eq:aInt2}
\end{equation}
where $A\sim \frac{T_{e}T_{p}v}{\omega_{p}-\omega_{e}}$ and $B\sim T_{p}^2$ are two constants. Eq.~(\ref{eq:aInt2}) is the same as Eq.~(\ref{eq:Int}) used in the Main Text to fit the experimentally-measured temperature-dependence of the integrated intensity of the A$_{1g}$ and E$_{g}^{(2)}$ phonon modes.

\vspace{-1mm}

%

\end{document}